\documentclass[aps,pre,twocolumn,groupedaddress]{revtex4}

\usepackage{graphicx}
\usepackage{bm}
\usepackage{amsmath}
\usepackage{subfigure}
\begin{document}

\title{Statistical mechanics of bend flexoelectricity and the twist-bend phase in bent-core liquid crystals}
\author{Shaikh M. Shamid}
\author{Subas Dhakal}
\thanks{Current address: Department of Biomedical and Chemical Engineering, Syracuse University, Syracuse, New York 13244.}
\author{Jonathan V. Selinger}
\email{jselinge@kent.edu}
\affiliation{Liquid Crystal Institute, Kent State University, Kent, Ohio 44242}

\date{May 7, 2013}

\begin{abstract}
We develop a Landau theory for bend flexoelectricity in liquid crystals of bent-core molecules. In the nematic phase of the model, the bend flexoelectric coefficient  increases as we reduce the temperature toward the nematic to polar phase transition. At this critical point, there is a second order transition from high-temperature uniform nematic phase to low-temperature nonuniform polar phase composed of twist-bend or splay-bend deformations. To test the predictions of Landau theory, we perform Monte Carlo simulations to find the director and polarization configurations as functions of temperature, applied electric field, and interaction parameters.
\end{abstract}

\maketitle

\section{Introduction}

Bent-core liquid crystals exhibit a rich variety of phases with different types of orientational order. At high temperature they form an isotropic phase, with disorder in all of the molecular axes. As the temperature decreases they form a nematic phase, in which the long axes of the molecules tend to align up or down along a director $\mathbf{\hat n}$. The transverse orientations of the molecules can then have various types of order in the plane perpendicular to $\mathbf{\hat n}$: they might be disordered (leading to a uniaxial nematic phase), or have nematic order (leading to a biaxial nematic phase), or have polar order (leading to a net polarization). A classic argument of Meyer~\cite{meyer69} shows that polar order of the transverse directions couples to bend variations in the main director $\mathbf{\hat n}$. As a result, it is particularly easy to induce polar order and director bend in bent-core liquid crystals, compared with analogous rod-like liquid crystals.

Recent research has found two remarkable physical phenomena arising from polar order and director bend in bent-core liquid crystals:

\paragraph{Flexoelectricity:} The flexoelectric effect is a linear coupling between polar order and director deformations in the uniaxial or biaxial nematic phase. An imposed deformation of the nematic director leads to an electrostatic polarization, and conversely, an applied electric field induces a deformation of the nematic director. This effect was initially discovered by Meyer~\cite{meyer69}, and since then has been investigated by many researchers~\cite{somoza91}. It has possible applications in devices such as display panels, actuators, micropower generators, and electro-mechanical transducers for sensing or energy-harvesting uses. In general, the director deformation might be either splay or bend, but one would expect the bend flexoelectric effect to be dominant in bent-core liquid crystals. Indeed, recent experiments by Harden \emph{et al.}~\cite{harden06,harden08} have found that bent-core liquid crystals have a surprisingly large bend flexoelectric coefficient, up to 35 nC/m, roughly three orders of magnitude larger than the typical value of 3--20 pC/m in rod-like liquid crystals.

For flexoelectricity, the key theoretical question is how to explain the large effect found in bent-core liquid crystals, so that it can be exploited for technological application. In a previous paper~\cite{subas10}, we conjectured that the large flexoelectric effect is a statistical phenomenon associated with nearby polar phase. Near a polar phase, a nematic liquid crystal is on the verge of developing spontaneous polar order, and hence any deformation of the director should induce a large polar response. In the previous paper, we explored that concept by constructing a model for splay flexoelectricity in a system of pear-shaped molecules. It is still necessary to extend the model to the more complex but physically realistic case of bend flexoelectricity in bent-core liquid crystals.

We recognize that experimental measurements of flexoelectricity in bent-core liquid crystals are controversial~\cite{coles2011,palffy2013}.  Regardless of the experimental flexoelectric coefficient in any particular material, we would like to understand what behavior is theoretically possible.

\paragraph{Modulated phases:} In an influential theoretical paper~\cite{dozov2001}, Dozov noted that bent-core liquid crystals have a very low energy cost for bend in the director, and described this low energy cost by a very small bend elastic constant $K_3$. He then speculated that $K_3$ could actually become \emph{negative} in some bent-core liquid crystals. (In this case, the free energy would need to be stabilized by higher-order terms in the bend.) Dozov showed theoretically that the uniform nematic phase would become unstable to the formation of a modulated phase, which could be either a \emph{twist-bend} or a \emph{splay-bend} phase.  Related simulations have been done by Memmer~\cite{memmer2002}.  This theoretical prediction is now supported experimentally by optical observations of spontaneous periodic deformations in bimesogens~\cite{panov2010}, and by extensive studies of a liquid crystal dimer, which identify the twist-bend phase using a series of techniques including small-angle x-ray scattering, modulated differential scanning calorimetry, dielectric spectroscopy, and magnetic resonance~\cite{luckhurst2011}.

For the twist-bend and splay-bend modulated phases, there are two important theoretical questions. First, how does the elastic constant $K_3$ become negative; what is the physical meaning of this negative value?  Second, what happens to the polar order in these phases? The schematic illustrations in Dozov's paper~\cite{dozov2001} clearly show the presence of local polar order, but this polar order is not included in his theoretical formalism.

The purpose of this paper is to develop a theory of polar order and director bend in bent-core liquid crystals, which can explain both flexoelectricity and modulated phases in these materials. For this work, we use both Landau theory and lattice simulations. In Landau theory, we construct a free energy functional in terms of director gradients and polarization. In the uniform nematic phase, we can minimize the free energy to find the optimal polarization for fixed bend, or conversely the optimal bend for fixed polarization (or applied electric field). This minimization gives the bend flexoelectric coefficient $e_3$, as well as the effective bend elastic constant and effective dielectric coefficient. In particular, it shows the difference between the bare elastic constant $K_3$ and the effective (or renormalized) elastic constant $K_3^\mathrm{eff}$. As the temperature decreases toward a critical temperature, $e_3$ increases and $K_3^\mathrm{eff}$ goes to zero. Below that critical temperature, the uniform nematic phase becomes unstable to the formation of a new phase which has both polar order and spontaneous bend, and hence must have a modulated structure. By minimizing the free energy over variational forms for the director and the polarization, we find that the director configuration is equivalent to Dozov's twist-bend or splay-bend phase, and we also determine the accompanying polarization configuration.

In simulations, we construct a lattice Hamiltonian that generalizes the classic Lebwohl-Lasher model~\cite{lebwohl72} for nematic liquid crystals by including two vectors on each site, which represent the long molecular axis and the transverse direction, respectively. This model is analogous to our previous lattice model for splay flexoelectricity~\cite{subas10}, but now extended to bend flexoelectricity. We run Monte Carlo simulations of this Hamiltonian in both the high-temperature uniform nematic phase and the low-temperature modulated polar phase. These simulations give results that are equivalent to Landau theory, but without the mean-field and variational assumptions of Landau theory. In the uniform nematic phase, they show the increasing flexoelectric coefficient as the temperature decreases. In the low temperature phase, they show the structure of the modulated phase, which may be twist-bend or splay-bend, depending on model parameters.

The plan of this paper is as follows. In Secs.~II and~III, we develop a theory for the flexoelectric effect in the uniform nematic phase, first using Landau theory and then lattice simulations. In Secs.~IV and~V, we extend the theory to the low-temperature modulated polar phase, now in the reverse sequence of lattice simulations and then Landau theory. In Sec.~VI we discuss and summarize the conclusions of this study.  Finally, in the Appendix we show how the lattice model can also be solved using mean-field theory.

\section{Flexoelectric effect: Landau theory}

To describe the flexoelectric effect in a uniform nematic phase, we must construct the free energy density in terms of the nematic director $\mathbf{\hat n}$ and the polarization $\mathbf{P}$. We assume that gradients of $\mathbf{\hat n}$ are small, and $\mathbf{P}$ is also small, so that we can work to quadratic order in both of these quantities.  (We will go to higher order in a later section, when discussing modulated phases.)  The free energy has three parts:

a.~The free energy of director gradients is just the standard Oseen-Frank free energy. It can be written as
\begin{equation}
F_{nn}= \frac{1}{2}K_{1}\mathbf{S}^2+\frac{1}{2}K_{2}\mathcal{T}^2+ \frac{1}{2}K_{3}\mathbf{B}^2,
\end{equation}
in terms of the splay vector $\mathbf{S}=\mathbf{\hat n}(\nabla\cdot\mathbf{\hat n})$, the twist pseudoscalar $\mathcal{T}=\mathbf{\hat n}\cdot(\nabla\times\mathbf{\hat n})$, and the bend vector $\mathbf{B}=\mathbf{\hat n}\times(\nabla\times\mathbf{\hat n})$.

b.~Polar order does not occur in the uniform nematic phase, and hence it must have a positive free energy cost. To lowest order, this cost can be written as
\begin{equation}
F_{pp}=\frac{1}{2}\mu\mathbf{P}^2,
\end{equation}
where $\mu$ is an arbitrary coefficient. There may be several physical contributions to $\mu$. Entropy makes a positive contribution proportional to temperature $T$, and the electrostatic energy makes another positive contribution independent of $T$. Because we are modeling bent-core liquid crystals, we suppose that there are packing considerations that \emph{favor} polar order, in competition with entropy and electrostatics. This packing energy can be modeled by a \emph{negative} part of the free energy, proportional to $\mathbf{P}^2$, independent of $T$. Hence, the combination of these effects gives a coefficient $\mu$ that varies with temperature, and can be written as $\mu=\mu'(T-T_0)$. Note that $T-T_0$ is positive in the uniform nematic phase.

c.~Following the argument of Meyer~\cite{meyer69}, polar order is coupled to both splay and bend in the director. We already considered the splay coupling in our previous paper on splay flexoelectricity~\cite{subas10}. In this paper we are concerned with bend flexoelectricity, and hence we consider the coupling
\begin{equation}
F_{np}=-\lambda\mathbf{B}\cdot\mathbf{P},
\end{equation}
which favors polar order along the bend direction, perpendicular to the director, and hence describes ordering of the transverse orientations of the molecules.

Putting these three pieces together, we obtain the total free energy density
\begin{equation}
F= \frac{1}{2}K_{1}\mathbf{S}^2+\frac{1}{2}K_{2}\mathcal{T}^2+ \frac{1}{2}K_{3}\mathbf{B}^2+\frac{1}{2}\mu\mathbf{P}^2-\lambda\mathbf{B}\cdot\mathbf{P}
\label{fquadratic}
\end{equation}
to quadratic order in director gradients and polar order. Note that the last three terms are a quadratic form in $\mathbf{B}$ and $\mathbf{P}$. The uniform nematic phase is only stable if the quadratic form is positive-definite, which occurs if $\lambda^2 < \mu K_3$. This condition can be rewritten as
\begin{equation}
\mu > \mu_c = \frac{\lambda^2}{K_3},
\end{equation}
or equivalently as
\begin{equation}
T > T_c = T_0 + \frac{\lambda^2}{\mu' K_3}.
\label{tc}
\end{equation}
Below that critical temperature, the uniform nematic phase must become unstable to a phase with director gradients and polar order, which will be discussed in Secs.~IV and~V below.

To model the flexoelectric effect, we minimize the free energy of Eq.~(\ref{fquadratic}) over the polarization $\mathbf{P}$ for fixed bend $\mathbf{B}$ to obtain
\begin{equation}
\textbf{P}=e_{3}\textbf{B},
\label{polarization}
\end{equation}
where the bend flexoelectric coefficient is given by
\begin{equation}
e_{3}=\frac{\lambda}{\mu}=\frac{\lambda}{\mu'(T-T_0)}.
\end{equation}
This coefficient increases toward a finite limit as the temperature decreases toward the critical temperature $T_c$. Substituting the expression~(\ref{polarization}) for $\textbf{P}$ back into Eq.~(\ref{fquadratic}) gives the effective free energy
\begin{equation}
F^\mathrm{eff}=\frac{1}{2}K_{1}\mathbf{S}^2+\frac{1}{2}K_{2}\mathcal{T}^2+ \frac{1}{2}K_{3}^\mathrm{eff}\mathbf{B}^2,
\end{equation}
where the effective, renormalized bend elastic constant is given by
\begin{equation}
K_{3}^\mathrm{eff}=K_{3}-\frac{\lambda^2}{\mu}=K_{3}-\frac{\lambda^2}{\mu'(T-T_0)}.
\label{K3eff}
\end{equation}
Note that there is an important physical distinction between the bare elastic coefficient $K_{3}$ and the renormalized coefficient $K_{3}^\mathrm{eff}$: the bare coefficient $K_{3}$ gives the energy cost of a bend in a hypothetical experiment where the polarization is constrained to be zero, while the renormalized coefficient $K_{3}^\mathrm{eff}$ gives the energy cost of a bend in an experiment where the polarization is free to relax to its lowest-free-energy value.  This distinction is essentially the same as the distinction between the bend coefficient $K_3^D$ at constant electric displacement $\mathbf{D}$ compared with $K_3^E$ at constant electric field $\mathbf{E}$, as discussed by Castles \emph{et al.}\ in the context of blue phase stability~\cite{castles2010}.  The renormalized coefficient is apparently the coefficient $K_{3}^\mathrm{eff}$ calculated by Cestari \emph{et al.}\ through molecular field theory with atomistic modeling, because their calculation involves averaging over molecular distributions that respond to bend~\cite{cestari2011}.  The bare coefficient $K_{3}$ is always positive, while the renormalized coefficient $K_{3}^\mathrm{eff}$ passes through zero as the temperature passes through $T_c$.  In Sec. V below, we will discuss the behavior below $T_c$, and will show that $K_{3}^\mathrm{eff}$ is the negative bend constant discussed by Dozov~\cite{dozov2001}.

To model the \emph{converse} flexoelectric effect, we add an electric field coupling to the polarization into the free energy~(\ref{fquadratic}), which gives
\begin{equation}
F= \frac{1}{2}K_{1}\mathbf{S}^2+\frac{1}{2}K_{2}\mathcal{T}^2+ \frac{1}{2}K_{3}\mathbf{B}^2
+\frac{1}{2}\mu\mathbf{P}^2-\lambda\mathbf{B}\cdot\mathbf{P}-\mathbf{E}\cdot\mathbf{P}.
\end{equation}
We minimize this free energy over \emph{both} bend $\mathbf{B}$ and polarization $\mathbf{P}$ for fixed electric field $\mathbf{E}$ to obtain
\begin{subequations}
\begin{align}
\mathbf{B} & = \frac{\lambda\mathbf{E}}{\mu K_3-\lambda^2},
\label{Beq}\\
\mathbf{P} & = \frac{K_3\mathbf{E}}{\mu K_3-\lambda^2}.
\label{Peq}
\end{align}
\end{subequations}
Equation~(\ref{Beq}) shows that an electric field induces a bend through the converse flexoelectric effect, while Eq.~(\ref{Peq}) shows that an electric field induces a polarization. The latter equation can be compared with the standard expression for the induced polarization in a dielectric material, $\mathbf{P}=\epsilon_{0} (\epsilon-1)\mathbf{E}$. This comparison shows that the effective dielectric constant is
\begin{equation}
\epsilon^\mathrm{eff}=1+\frac{1}{\epsilon_{0}(\mu-\lambda^2/K_{3})}.
\end{equation}
Hence, the effective dielectric constant is renormalized upward by the coupling between polarization and bend, compared with the value $\epsilon=1 + 1/(\epsilon_{0}\mu)$ without this coupling.  At the critical temperature $T_c$ where the uniform nematic phase becomes unstable, the effective dielectric constant diverges.

Here, we should emphasize which quantities diverge and which remain finite at the critical point $T_c$. When we apply an electric field in the converse flexoelectric effect, the induced bend and polarization are given by susceptibilities multiplying the field.  These susceptibilities both diverge as $T\to T_c$, as usual for susceptibilities at a second-order phase transition.  Likewise, if we were to apply a torque that couples directly to the bend, the induced bend and polarization would be given by susceptibilities multiplying the torque, and those susceptibilities would also diverge as $T\to T_c$.  However, the flexoelectric coefficient $e_3$ is not exactly a susceptibility; rather, it is the \emph{ratio} between bend and polarization, which both diverge at the transition.  This ratio does not diverge, but approaches the finite maximum value $e_3^\mathrm{max}=K_3/\lambda$.  By comparing this maximum value with the bend elastic constant and dielectric constant  as $T\to T_c$, we see that
\begin{equation}
(e_3^\mathrm{max})^2 = K_3 \epsilon_0 (\epsilon-1) = K_3^\mathrm{eff} \epsilon_0 (\epsilon^\mathrm{eff}-1).
\end{equation}
This equation for the maximum flexoelectric coefficient is approximately the same as the limit derived by Castles \emph{et al.}~\cite{coles2011}.  However, they derive the limit based on arguments about the conservation of energy.  We would not say that it is related to energy conservation, but rather that it is the limit of stability of the uniform nematic phase.  Beyond that point, the uniform nematic phase becomes unstable to the formation of a modulated phase, as will be discussed in later sections.

\section{Flexoelectric Effect:  Lattice Model}

In the previous section, we developed a Landau theory for the bend flexoelectric effect near a transition to an incipient polar phase.  In this section, we develop a lattice simulation model to describe the same effect.  This model is an extension of the lattice model for splay flexoelectricity in our previous paper~\cite{subas10}.  The lattice simulations will provide clear visualizations of the types of molecular order that occur in the flexoelectric effect.  They will also allow us to avoid the standard limitations of Landau theory, which neglects correlated fluctuations and applies only over a limited temperature interval.

\begin{figure}
\includegraphics[width=3.375in]{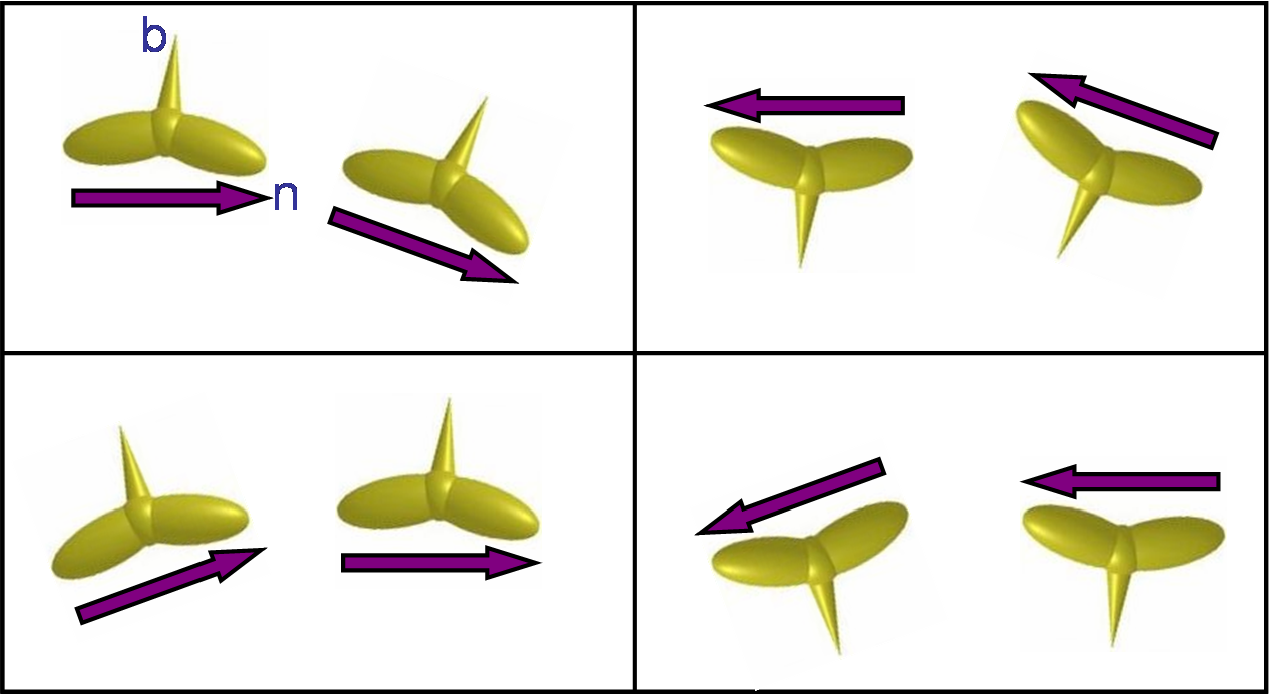}
\caption{\label{Fig1} (Color online) Lattice model for neighboring bent-core liquid crystal molecules, showing possible orientations of two neighboring molecules under a bend deformation.}
\end{figure}

We consider a simple cubic lattice with a bent-core liquid crystal molecule on each site. The orientation of each molecule is characterized by two orthogonal unit vectors, as shown in Fig.~\ref{Fig1}.  The vector $\hat{\bm{n}}_i$ represents the long molecular axis, while the vector $\hat{\bm{b}}_i$ represents the transverse dipole of the molecule at site $i$.  In the uniaxial nematic phase \emph{without} any bend, the $\hat{\bm{n}}$ vectors are ordered but the $\hat{\bm{b}}$ vectors are disordered.  We would like to model how a bend in $\hat{\bm{n}}$  induces polar order in $\hat{\bm{b}}$, or conversely, how an electric field applied to $\hat{\bm{b}}$ induces bend in $\hat{\bm{n}}$.

For the lattice interaction between molecules on neighboring sites $i$ and $j$, we must consider several terms.  First, we need the standard Lebwohl-Lasher~\cite{lebwohl72} interaction $-A(\hat{\bm{n}}_i\cdot\hat{\bm{n}}_j)^2$, which favors nematic order of the molecules.  Second, to describe a bent-core liquid crystal with incipient polar order, we need an interaction of the form $-B_1\hat{\bm{b}}_i\cdot\hat{\bm{b}}_j$.  This term favors polar order of the transverse dipoles, driven by packing energy or any other mechanism.  Third, there may also be a higher-order interaction of the form $-B_2(\hat{\bm{b}}_i\cdot\hat{\bm{b}}_j)^2$, which favors nematic order of the transverse dipoles, perpendicular to the main nematic axis.  This term gives the possibility of a biaxial nematic phase, as in the model of Straley~\cite{straley74}.  Fourth, there must be an interaction $-\bm{E}\cdot\hat{\bm{b}}_i$ between each dipole and the applied electric field.

Finally, we need a coupling of $\hat{\bm{b}}$ with the local bend in $\hat{\bm{n}}$.  Following the argument of Ref.~\cite{subas10}, a lattice version of the bend vector between sites $i$ and $j$ can be written as
\begin{eqnarray}
[\mathbf{\hat n}\times(\mathbf{\nabla}\times\mathbf{\hat n})]_{ij}&=&\frac{1}{2}\Bigl[
\hat{\bm{n}}_i(\hat{\bm{r}}_{ij}\cdot\hat{\bm{n}}_i)
-\hat{\bm{n}}_j(\hat{\bm{r}}_{ij}\cdot\hat{\bm{n}}_j)\nonumber\\
&&\quad+\hat{\bm{n}}_i(\hat{\bm{n}}_i\cdot\hat{\bm{n}}_j)(\hat{\bm{r}}_{ij}\cdot\hat{\bm{n}}_j)\label{bend}\\
&&\quad-\hat{\bm{n}}_j(\hat{\bm{n}}_i\cdot\hat{\bm{n}}_j)(\hat{\bm{r}}_{ij}\cdot\hat{\bm{n}}_i)\Bigr],\nonumber
\end{eqnarray}
where $\hat{\bm{r}}_{ij}=(\bm{r}_j-\bm{r}_i)/|\bm{r}_j-\bm{r}_i|$ is the unit vector from site $i$ to $j$ on the lattice.  Note that this expression is invariant under the symmetry operations $i \leftrightarrow j$, $\hat{\bm{n}}_i \rightarrow -\hat{\bm{n}}_i$, and $\hat{\bm{n}}_j\rightarrow-\hat{\bm{n}}_j$. This expression for the local bend can be coupled with the local polar order, averaged over sites $i$ and $j$, to give
\begin{eqnarray}
V_\mathrm{bend}&=&C[\mathbf{\hat n}\times(\mathbf{\nabla}\times\mathbf{\hat n})]_{ij}\cdot\frac{\hat{\bm{b}}_i+\hat{\bm{b}}_j}{2}\nonumber\\
&=&\frac{C}{4}\Bigl[(\hat{\bm{b}}_j\cdot\hat{\bm{n}}_i)[\hat{\bm{r}}_{ij}\cdot\{\hat{\bm{n}}_i
+\hat{\bm{n}}_j(\hat{\bm{n}}_i\cdot\hat{\bm{n}}_j)\}] \nonumber\\
&&-(\hat{\bm{b}}_i\cdot\hat{\bm{n}}_j)[\hat{\bm{r}}_{ij}\cdot\{\hat{\bm{n}}_j
+\hat{\bm{n}}_i(\hat{\bm{n}}_i\cdot\hat{\bm{n}}_j)\}]\Bigr].
\end{eqnarray}
The coefficient $C$ may be either positive or negative.  The sign of $C$ determines whether the coupling favors parallel or antiparallel alignment of polar order and bend, but does not otherwise affect the behavior.

Combining all the terms, our final expression for the lattice Hamiltonian is
\begin{eqnarray}
H&=&-\sum_{\langle i,j\rangle}\biggl[
A(\hat{\bm{n}}_i\cdot\hat{\bm{n}}_j)^2
+B_1\hat{\bm{b}}_i\cdot\hat{\bm{b}}_j
+B_2(\hat{\bm{b}}_i\cdot\hat{\bm{b}}_j)^2\nonumber\\
&&\qquad
-\frac{C}{4}\Bigl[(\hat{\bm{b}}_j\cdot\hat{\bm{n}}_i)[\hat{\bm{r}}_{ij}\cdot\{\hat{\bm{n}}_i
+\hat{\bm{n}}_j(\hat{\bm{n}}_i\cdot\hat{\bm{n}}_j)\}] \nonumber\\
&&\qquad\qquad
-(\hat{\bm{b}}_i\cdot\hat{\bm{n}}_j)[\hat{\bm{r}}_{ij}\cdot\{\hat{\bm{n}}_j
+\hat{\bm{n}}_i(\hat{\bm{n}}_i\cdot\hat{\bm{n}}_j)\}]\Bigr]\biggr]\nonumber\\
&&-\sum_i \bm{E}\cdot\hat{\bm{b}}_i.
\label{hamiltonian}
\end{eqnarray}
This expression is analogous to the lattice Hamiltonian in Ref.~\cite{subas10}, except that it has two orthogonal unit vectors on each site instead of just one, so that it can describe bend instead of splay flexoelectricity.

We carry out Monte Carlo simulations of a system of bent-core molecules interacting with the Hamiltonian of Eq.~(\ref{hamiltonian}) on a simple cubic lattice of size $16\times16\times16$.  When an electric field is applied, it is in the $Z$-direction. The simulation box has periodic boundary conditions in $Z$, but the boundaries in $X$ and $Y$ are free so that the system can form bend across those directions.  In each Monte Carlo step, a molecule is chosen randomly and it is slightly rotated about a random axis. The change in energy $\Delta E$ is calculated, and the step is accepted or rejected following the Metropolis algorithm.  Starting from a high-temperature isotropic state, the system is cooled down  slowly with temperature steps of  $\Delta T=0.01$. The final configuration at each temperature is taken as the initial configuration for the next lower temperature. Typical runs take about $10^5$ steps to come to equilibrium, while runs near phase transitions take about $6\times 10^5$ steps.

\begin{figure}
\includegraphics[width=3.375in]{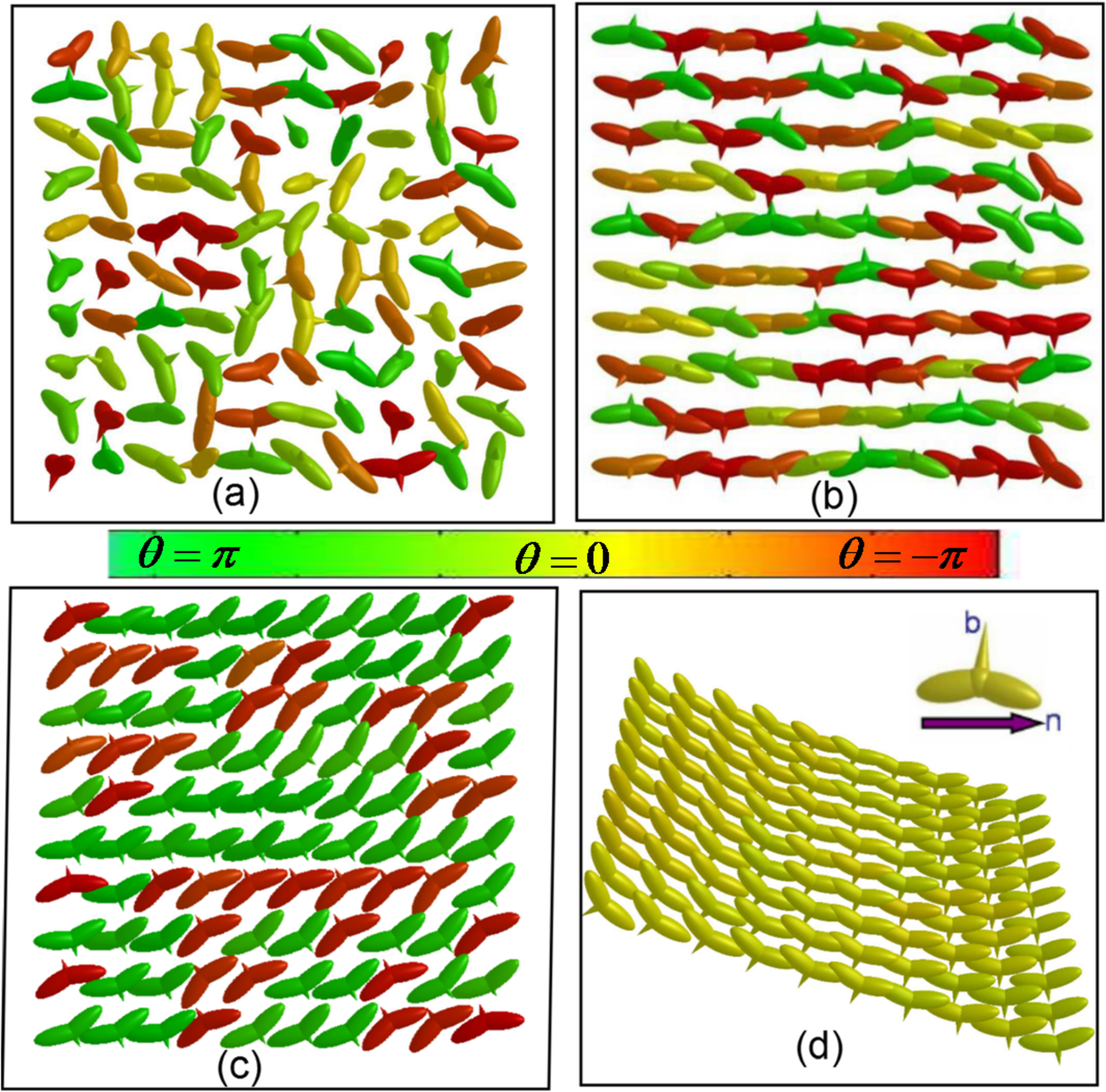}
\caption{\label{Fig2}~(Color online) Equilibrium configurations from Monte Carlo simulations showing (a)~isotropic phase,  (b)~uniaxial nematic phase, (c)~biaxial nematic phase, and (d)~polar phase. The color on each molecule represents the orientation ($\theta$) of its short axis $\hat{\bm{b}}$ with respect to the laboratory $Z$-axis.}
\end{figure}

Figure~\ref{Fig2} shows snapshots of the molecular configuration in four phases at zero electric field.  At high temperature, the system is in an isotropic ($I$) phase, with disorder in both $\hat{\bm{n}}$ and $\hat{\bm{b}}$.  As the temperature decreases, it forms a uniaxial nematic ($N_U$) phase, with nematic order in $\hat{\bm{n}}$ but disorder in $\hat{\bm{b}}$ vectors, which are uniformly distributed in the plane perpendicular to the average director.  At lower temperature, it forms a biaxial nematic ($N_B$) phase, with nematic order in both $\hat{\bm{n}}$ and $\hat{\bm{b}}$.  In the $N_B$ phase, the $\hat{\bm{b}}$ vectors have two favored orientations perpendicular to the average director.  At the lowest temperature, it forms a polar ($P$) phase, with nematic order in $\hat{\bm{n}}$ and polar order in $\hat{\bm{b}}$ vectors, which now have one favored orientation perpendicular to the average director.  In this case, the the polar order induces bend in the director.  (At longer length scales more complex modulated structures are seen, as discussed in the following section.)

To characterize the orientational order of these phases, we use a standard set of four uniaxial and biaxial nematic order parameters, supplemented by an additional polar order parameter.  The nematic order is represented by the supermatrix~\cite{allen90,philip99,luckhurst05}
\begin{equation}
S_{ij}^{IJ}= \left\langle \frac{3 l_{iI} l_{jJ}-\delta_{iI}\delta_{jJ}}{2} \right\rangle.
\end{equation}
Here, the subscripts $i$ and $j$ denote molecular axes while $I$ and $J$ represent laboratory axes, and $l_{iI}$ is the direction cosine between molecular and laboratory axes, $l_{iI}=\hat{\bm{i}}\cdot\hat{\bm{I}}$.  The reference frame, which we denote as $(\hat{\bm{N}},\hat{\bm{B}},\hat{\bm{C}})$, is calculated as follows~\cite{allen90,philip99,luckhurst05}: First, the $Q$-tensor is calculated for all molecular axes $(\hat{\bm{n}},\hat{\bm{b}},\hat{\bm{c}})$. The eigenvector corresponding to the largest eigenvalue is the reference $N$ axis. Similarly, the second largest eigenvalue is taken to identify the secondary molecular ordering axis. The corresponding eigenvector is the reference $B$ axis. The remaining reference axis $C$ is perpendicular to $N$ and $B$. In this frame, the nematic order parameters can be expressed as
\begin{eqnarray}
S &=& S_{nn}^{NN}, \quad U = S_{bb}^{NN} - S_{cc}^{NN}, \quad \mathsf{T} = S_{nn}^{BB} - S_{nn}^{CC},\nonumber\\
V &=& \frac{1}{3}[(S_{bb}^{BB} - S_{bb}^{CC}) - (S_{cc}^{BB} - S_{cc}^{CC})].
\end{eqnarray}
As a physical interpretation, $S$ measures the ordering of the molecular $n$-axis with respect to the $N$-axis and is the usual nematic order parameter, while $U$ measures the difference in ordering of the molecular $b$ and $c$-axes with respect to the $N$-axis.  Conversely, $\mathsf{T}$ measures the difference in ordering of the molecular $n$-axis with respect to the reference $B$ and $C$-axes. Finally, $V$ measures the ordering of the molecular $b$ and $c$-axes with respect to the reference $B$ and $C$-axes. Of these four parameters, $S$ and $U$ are non-zero in the uniaxial and biaxial phases, while $\mathsf{T}$ and $V$ vanish in the uniaxial phase but are non-zero in the biaxial phase. In the limit of low temperature and high orientational order, $S$ and $V$ tend to $1$, while $U$ and $\mathsf{T}$ tend to $0$.  We calculate all four of these order parameters following the procedure of Refs.~\cite{allen90,philip99,luckhurst05}, but we only present results for the uniaxial nematic order parameter $S$ and biaxial nematic order parameter $V$.


In addition to the nematic order parameters, we define the polar order parameter as the average of the molecular dipole vectors, $\bm{P}=\langle\hat{\bm{b}}\rangle$, and we present results for the magnitude $P=|\bm{P}|$.

\begin{figure}
\includegraphics[width=3.375in]{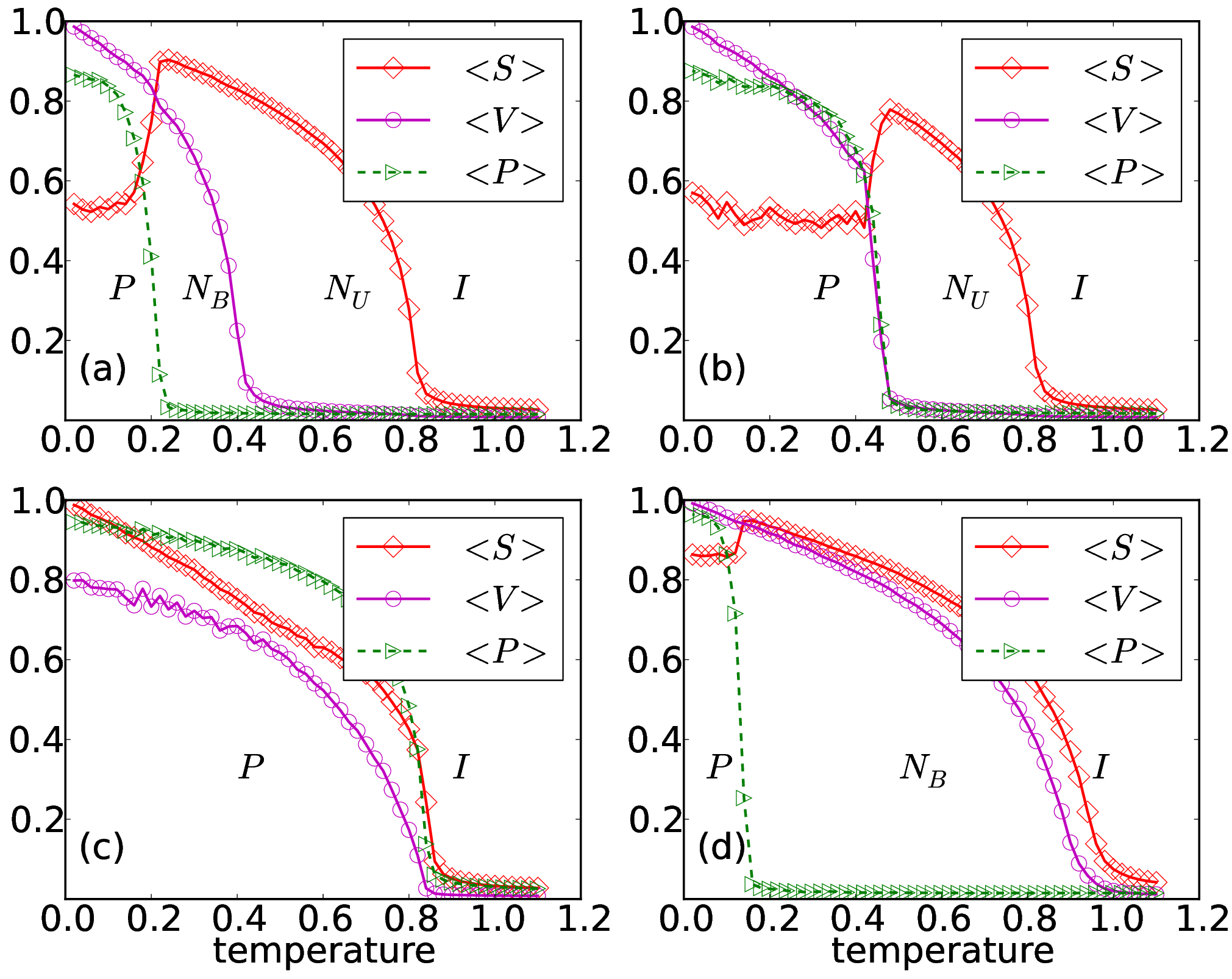}
\caption{\label{Fig3}~(Color online) Monte Carlo simulation results for the order parameters $S$~($\triangleright$), $V$~($\times$) and $P$~($\circ$) as functions of temperature $T$, for different values of the coupling coefficients chosen to show different types of transitions.  (a)~$B_1=0.04$, $B_2=0.3$.  (b)~$B_1=0.12$, $B_2=0.3$.  (c)~$B_1=0.4$, $B_2=0.3$.  (d)~$B_1=0.04$, $B_2=0.95$.  In all cases $A=1.0$, $C=0.3$, and $E=0$.}
\end{figure}

The order parameters $S$, $V$, and $P$ are time-averaged in the production cycle of simulations.  Plots of the order parameters for different values of the coupling coefficients are shown in Fig.~\ref{Fig3}.  In all these cases, the isotropic phase is stable at high temperatures, followed by phases of reduced symmetry at low temperatures. We find four types of transitions depending upon the relative coupling coefficients:  (i)~For small biaxial coupling $B_2$ and even smaller polar coupling $B_1$, there is a series of transitions $I\rightarrow N_U\rightarrow N_B\rightarrow P$. At each transition, the system acquires additional order. (ii)~For somewhat larger polar coupling $B_1$, a direct transition from uniaxial nematic to polar phase occurs, with no intermediate biaxial nematic phase, $I\rightarrow N_U\rightarrow P$. (iii)~For even stronger $B_1$, a direct transition from isotropic to polar phase takes place, with no intermediate uniaxial or biaxial nematic phase, $I\rightarrow P$. (iv) If the molecules have a very strong biaxial coupling $B_2$, the transition goes directly from isotropic to biaxial nematic, with no intermediate uniaxial nematic phase, $I\rightarrow N_B\rightarrow P$. From these observations, we see that the stability of the biaxial nematic phase can be tuned with the polar strength~$B_1$ and biaxiality~$B_2$ of the molecules.

\begin{figure}
\includegraphics[width=3.375in]{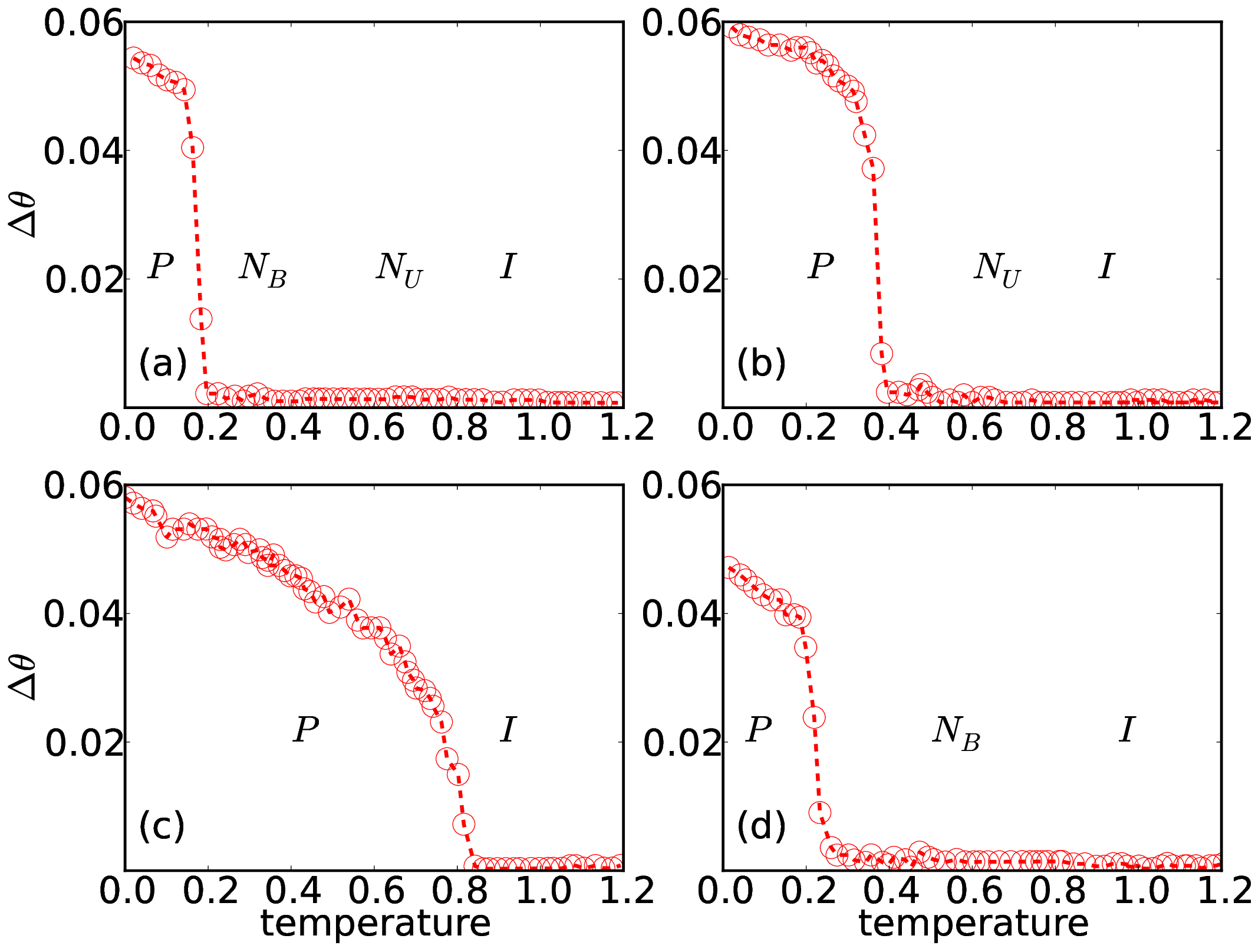}
\caption{\label{Fig4}~(Color online) Monte Carlo simulation results for the bend per bond as a function of temperature $T$, for different values of the coupling coefficients.  (a)~$B_1=0.04$, $B_2=0.3$.  (b)~$B_1=0.12$, $B_2=0.3$.  (c)~$B_1=0.4$, $B_2=0.3$.  (d)~$B_1=0.04$, $B_2=0.95$.  In all cases $A=1.0$, $C=0.3$, and $E=0$.}
\end{figure}

A key feature of our model is that bend in $\hat{\bm{n}}$ is coupled to polar order in $\hat{\bm{b}}$.  For that reason, we expect nonzero bend to occur spontaneously as the system moves into the polar phase, even under zero applied electric field.  To test this concept, we calculate the average bend $\Delta\theta$, defined by averaging Eq.~(\ref{bend}) over bonds, as a function of temperature and coupling coefficients.  The results are shown in Fig.~\ref{Fig4}.  Note that this spontaneous bend is zero in the isotropic, uniaxial nematic, and biaxial nematic phases, and becomes nonzero in the polar phase.  One further point about the order parameters should be noted here:  We calculate $S$, $V$, and $P$ by averaging over the entire system.  In the polar phase, because there is a systematic bend in the molecular orientation, these global averages are reduced.  That is the reason why Fig.~\ref{Fig3} shows a decrease in $S$ in the polar phase.  Presumably this issue could be avoided by calculating \emph{local} averages of the order parameters, but we use global averages because of the limited system size.

\begin{figure}
\includegraphics[width=3.375in]{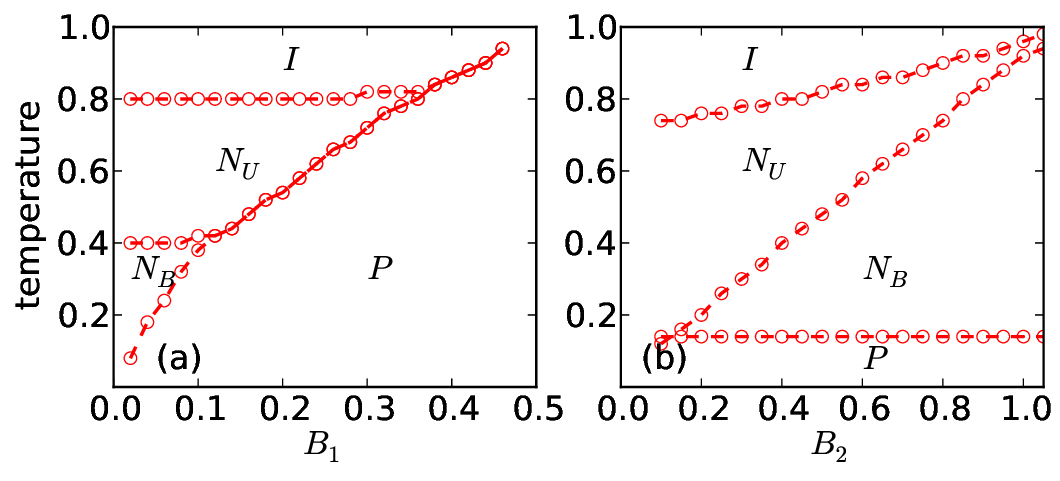}
\caption{\label{Fig5} (Color online) Phase diagram of the model system as a function of (a)~$T$ and $B_2$, for fixed $B_1=0.03$.  (b)~$T$ and $B_1$, for fixed $B_2=0.45$.  In all cases $A=1.0$, $C=0.3$, and $E=0$.}
\end{figure}

To explore the phase diagram of the system, we carry out a series of simulations where we vary the coupling coefficients $B_1$ and $B_2$ and measure the resulting order parameters.  Figure~\ref{Fig5} shows two cross-sections through the phase diagram.  Note that increasing $B_2$ enhances the stability of the biaxial nematic phase at the expense of the uniaxial nematic and isotropic phases, but does not affect the polar phase.  By comparison, increasing $B_1$ enhances the stability of the polar phase at the expense of all the other phases.

\begin{figure}
\includegraphics[width=3.375in]{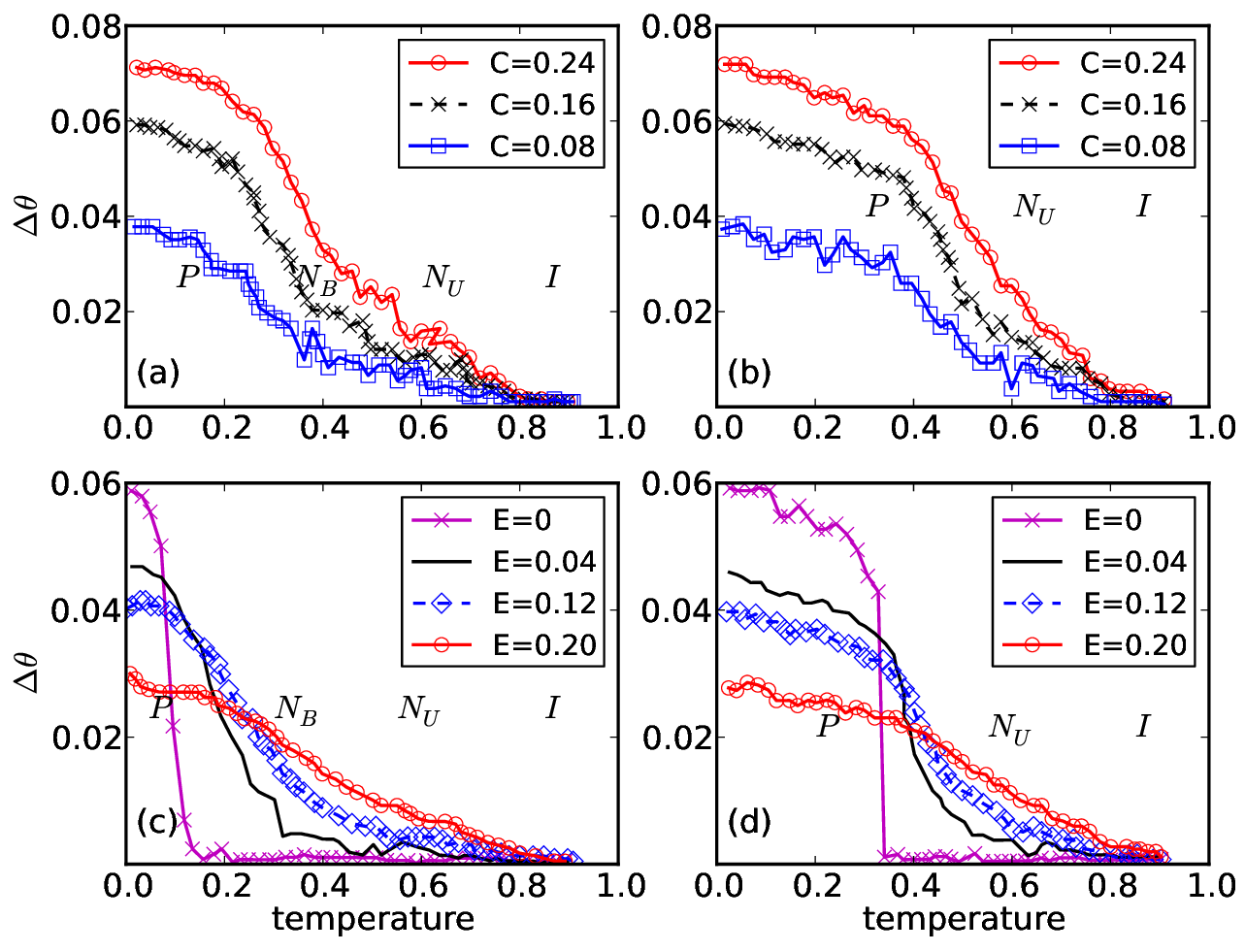}
\caption{\label{Fig6} (Color online) Bend as a function of temperature for the system under an electric field.  (a,~b)~For different values of the bend-polarization coupling $C$.  (c,~d)~For different values of the electric field $E$.  Parts (a) and (c) show systems with a biaxial nematic phase, with parameters as in Fig.~\ref{Fig3}a, while (b) and (d) show systems with a direct transition from uniaxial nematic to polar, with parameters as in Fig.~\ref{Fig3}b.}
\end{figure}

We now simulate the converse flexoelectric effect by applying an electric field $E$ along the $Z$-axis.  As expected, the dipole direction $\hat{\bm{b}}$ aligns parallel to the field, while the director $\hat{\bm{n}}$ bends across the system, as a function of $X$ or $Y$.  We calculate this induced bend from Eq.~(\ref{bend}) as a function of temperature, electric field, and coupling coefficients.  The results are shown in Fig.~\ref{Fig6}.  Note that the bend increases as the field increases, as the magnitude of bend-polarization coupling $C$ increases, and as the temperature decreases.  In particular, as the temperature decreases toward the transition into the polar phase, the bend responds sensitively to any applied field.  The temperature dependence of the bend is sharpest for very low fields, and it is rounded off for larger fields.  This behavior is similar regardless of whether the system passes through the biaxial nematic phase or goes directly from uniaxial nematic to polar.  These trends are the normal behavior of an induced order parameter in the presence of a symmetry-breaking field above a second-order phase transition.  They are consistent with the Landau theory presented in the previous section, which suggests that the flexoelectric and converse flexoelectric effects become large near an incipient polar phase, where the uniform nematic phase is almost unstable and the system is most sensitive to any symmetry-breaking perturbation.

\begin{figure}
\includegraphics[width=2.1in]{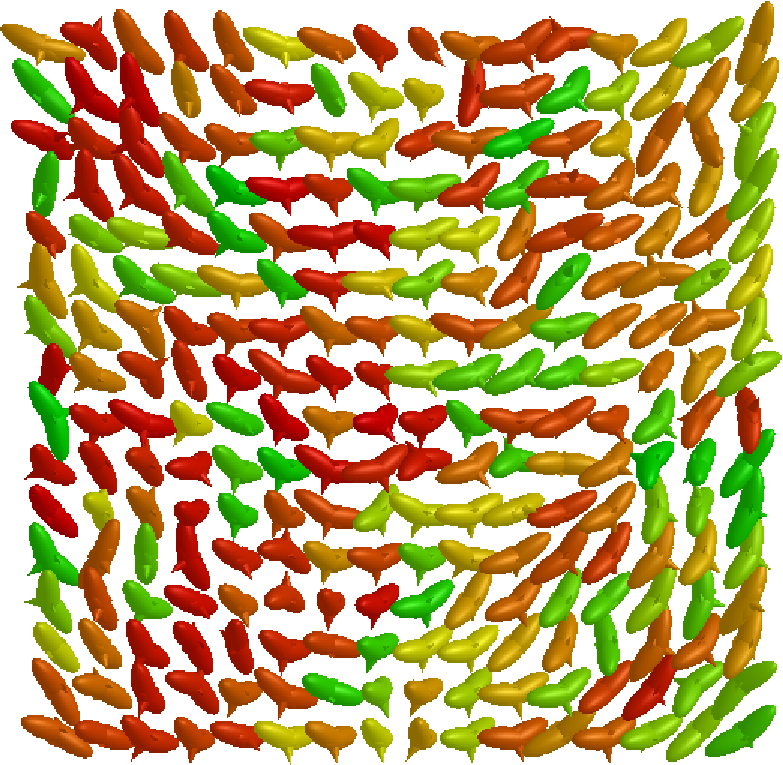}
\caption{\label{Fig7}~(Color online) Equilibrium configuration from a Monte Carlo simulation just above the uniaxial nematic-polar transition temperature for $E=0.04$.}
\end{figure}

To visualize the molecular configuration responsible for the large flexoelectric effect, we perform a simulation in the uniaxial nematic phase for $E=0.04$ at $T=0.42$, slightly above the transition into the polar phase at $T_{UP}\approx 0.32$.  A snapshot of the configuration is shown in Fig.~\ref{Fig7}.  Clearly there are local correlated regions with bend and polar order.  At this temperature, the system is highly susceptible to an applied electric field, which orders the local correlated regions and induces a global polarization and bend.  Presumably it would also be highly susceptible to an applied torque coupling to the bend, which would have the same effects, although we have not done that simulation explicitly.

Apart from the Monte Carlo simulations, the lattice model of Eq.~(\ref{hamiltonian}) can also be solved analytically through mean-field theory.  The mean-field calculation is presented in the Appendix, and the results are similar to the simulations presented in this section.

\section{Modulated phases:  Lattice model}

As noted in the previous section, when the system cools into a polar phase with spontaneous polar order, it also acquires spontaneous bend.  A simple example of the spontaneous bend is illustrated in Fig.~\ref{Fig2}d, which shows a gradual bend across the system, between the free boundaries.  However, this simple configuration cannot be extended to give the molecular orientation across a larger system.  In general, it is impossible to fill space with pure uniform bend.  Rather, the system must form a more complex modulated phase, which might have a regular array of defect walls, or might have a mixture of bend with splay or twist in the director.  Indeed the problem of filling space with bend in a polar liquid-crystal phase is quite analogous to the problem of filling space with twist in a chiral liquid-crystal phase, as discussed in Ref.~\cite{kamien01}.

To determine the modulated structure of the polar phase, we repeat the simulations of the previous section with three modifications:  we use periodic boundary conditions in all three directions, we use a slightly larger lattice of size $20\times 20 \times 20$, and we increase the magnitude of bend-polarization coupling $C$ to increase the bend, i.~e. reduce the wavelength of the modulated structure, so that a full wavelength will fit in the simulation box.  In these simulations no electric field is applied, so the only polarization is spontaneous order.  We begin the simulations in the high-temperature isotropic phase and gradually cool into the low-temperature polar phase.  In this way, the system is free to select its own modulated structure.

\begin{figure}
(a)\subfigure{\includegraphics[height=2.4in]{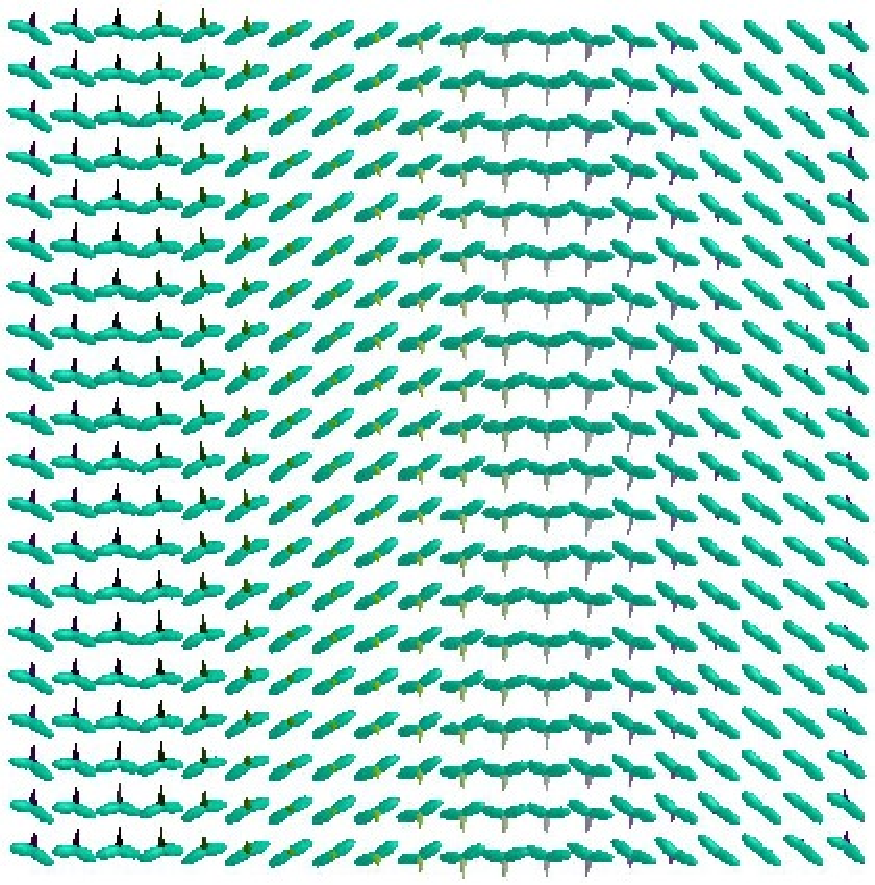}}
(b)\subfigure{\includegraphics[height=2.4in]{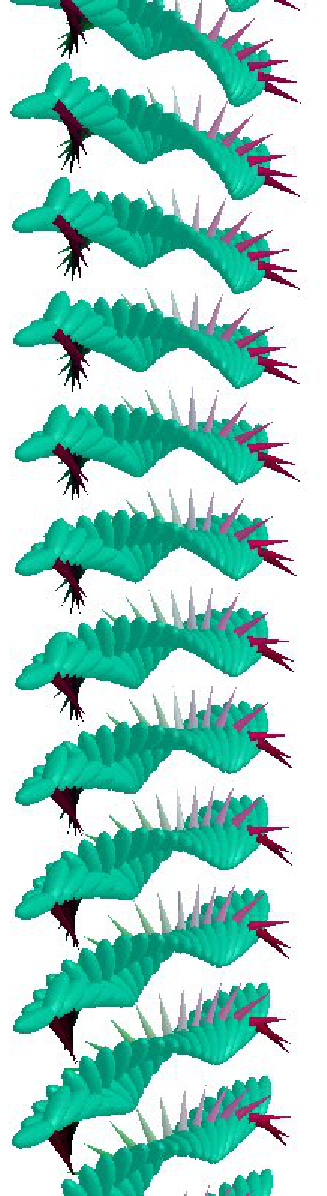}}
\caption{\label{Fig8}(Color online) Equilibrium configuration in a Monte Carlo simulation of the twist-bend phase, for model parameters $A=2$, $B_1=0.5$, $B_2=0.4$, $C=-2.0$, and $T=0.5$.  (a)~Top view.  (b)~Side view.}
\end{figure}

In these simulations, two distinct types of modulated structures form, depending on the model parameters.  The first structure, shown in Fig.~\ref{Fig8}, is equivalent to the twist-bend phase proposed by Dozov~\cite{dozov2001}.  In this structure, the director $\hat{\bm{n}}$ has a helical modulation, which is randomly right- or left-handed.  The director is not perpendicular to the helical axis, as in a cholesteric liquid crystal.  Rather, it precesses in a cone about the helical axis, with a fixed cone angle between $0^\circ$ and $90^\circ$.  As a result, the director deformation is a mixture of twist and bend, unlike a cholesteric liquid crystal which has pure twist.  The dipole direction $\hat{\bm{b}}$ \emph{also} precesses about the helical axis, while remaining perpendicular to $\hat{\bm{n}}$ \emph{and} perpendicular to the helical axis.  This structure is spatially homogeneous, in that every position is equivalent to every other position with a twist.  Hence, every position has the same energy, and there are no defects.

\begin{figure}
\includegraphics[width=3.0in]{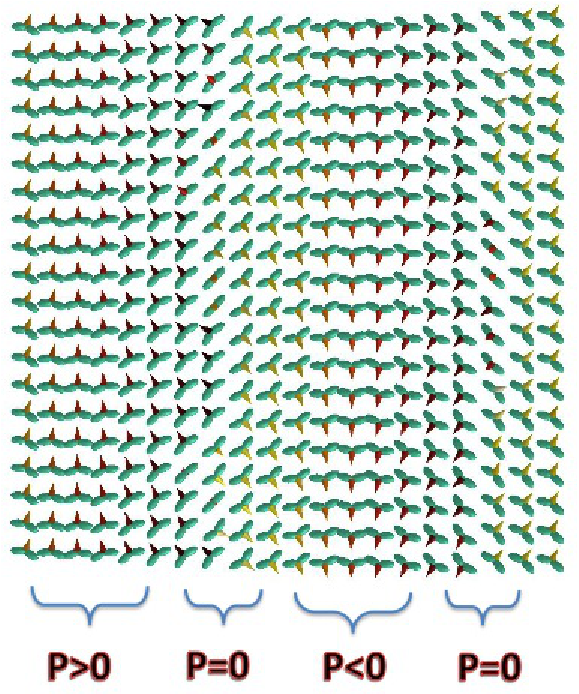}
\caption{\label{Fig9}(Color online) Equilibrium configuration in a Monte Carlo simulation of the splay-bend phase, for model parameters $A=3$, $B_1=0.15$, $B_2=0.4$, $C=-2.0$, and $T=0.3$.  The director and polar order are both in the plane of the figure.  The labels indicate the the local polarization along the vertical axis.}
\end{figure}

The second structure, shown in Fig.~\ref{Fig9}, is equivalent to the splay-bend phase proposed by Dozov~\cite{dozov2001}.  Here, the director $\hat{\bm{n}}$ goes back and forth within the plane of the figure.  As a result, the director deformation is a mixture of splay and bend.  Note that different regions are not equivalent to each other---some regions have almost pure bend, and other regions have almost zero bend.  In this structure, the local polar order varies in \emph{both} magnitude and direction.  In the pure-bend regions the $\hat{\bm{b}}$ vectors are very well aligned, and hence the polar order parameter has a large magnitude.  These regions are indicated by the labels $P>0$ and $P<0$ in the figure.  By contrast, in the zero-bend regions, the $\hat{\bm{b}}$ vectors are disordered, and hence the polar order parameter averages to zero.  These regions are indicated by the label $P=0$ in the figure.  The zero-bend regions have a higher energy than the pure-bend regions, so they can be regarded as defect walls.

The simple example of Fig.~\ref{Fig2}d can be understood as one pure-bend region going across the finite simulation cell.  The splay-bend structure of Fig.~\ref{Fig9} shows how this structure can fill space with a periodic alternation of pure-bend ``defect-free'' regions and zero-bend ``defect walls.''

When we say that the structures of Figs.~\ref{Fig8} and~\ref{Fig9} are equivalent to Dozov's twist-bend and splay-bend phases, we mean that the director modulations are the same as what he proposed.  His paper does not explicitly consider the polarization modulation, although his sketches suggest a variation in the polarization direction that is similar to our simulation results.  The twist-bend phase has also been visualized in a simulation by Memmer~\cite{memmer2002}.

\section{Modulated phases:  Landau theory}

To understand the modulated phases better, we return to the Landau theory of Sec.~II.  These phases occur for temperatures below the critical temperature of Eq.~(\ref{tc}), where the quadratic form in the free energy of Eq.~(\ref{fquadratic}) is not positive-definite.  Hence, we must add further terms to stabilize the free energy.  First, there must be a term of $\frac{1}{4}\nu\mathbf{P}^4$, which keeps the magnitude of the polar order from increasing without limit.  Second, there must be a term of $\frac{1}{2}\kappa(\nabla\mathbf{P})^2$, which penalizes gradients in the magnitude and direction of polar order.  (In terms of tensor indices, we interpret this gradient term as $\frac{1}{2}\kappa(\partial_i P_j)(\partial_i P_j)$.  There could be other tensor contractions, but they do not matter for our analysis.)  With these new terms, the Landau free energy becomes
\begin{eqnarray}
F&=& \frac{1}{2}K_{1}\mathbf{S}^2+\frac{1}{2}K_{2}\mathcal{T}^2+ \frac{1}{2}K_{3}\mathbf{B}^2-\lambda\mathbf{B}\cdot\mathbf{P}\nonumber\\
&&+\frac{1}{2}\mu'(T-T_0)\mathbf{P}^2+\frac{1}{4}\nu\mathbf{P}^4+\frac{1}{2}\kappa(\nabla\textbf{P})^2.
\label{fcomplete}
\end{eqnarray}

To model the twist-bend phase, we make the variational ansatz for $\mathbf{\hat n}(x)$ and $\mathbf{P}(x)$ inspired by the simulation results of Fig.~\ref{Fig8},
\begin{eqnarray}
\mathbf{\hat n}(x)&=&(1-a^2)^{1/2} \mathbf{\hat x}+a\sin(q x)\mathbf{\hat y}+a\cos(q x)\mathbf{\hat z},\nonumber\\
\mathbf{P}(x)&=&-p\cos(q x)\mathbf{\hat y}+p\sin(q x)\mathbf{\hat z}.
\end{eqnarray}
This ansatz has three variational parameters:  $a$ is the sine of the cone angle for the director, $p$ is the magnitude of the local polar order, and $q$ is the wavevector of the modulation.  In terms of these parameters, the splay, twist, and bend become
\begin{eqnarray}
\mathbf{S}(x)&=&\mathbf{\hat n}(\nabla\cdot\mathbf{\hat n})=0,\nonumber\\
\mathcal{T}(x)&=&\mathbf{\hat n}\cdot(\nabla\times\mathbf{\hat n})=a^2 q,\\
\mathbf{B}(x)&=&\mathbf{\hat n}\times(\nabla\times\mathbf{\hat n})\nonumber\\
&=&a(1-a^2)^{1/2} q[-\cos(q x)\mathbf{\hat y}+\sin(q x)\mathbf{\hat z}].\nonumber
\end{eqnarray}
Note that the splay is zero, as it should be for the twist-bend deformation.  The twist and the bend magnitude are constants, while the bend direction precesses in a helix.  Plugging those quantities into the free energy gives
\begin{eqnarray}
F_{TB}&=&\frac{1}{2}K_{2} a^4 q^2 +\frac{1}{2}K_{3} a^2(1-a^2)q^2 -\lambda a p q (1-a^2)^{1/2}\nonumber\\
&&+\frac{1}{2}\mu'(T-T_0) p^2 +\frac{1}{4}\nu p^4 +\frac{1}{2} \kappa p^2 q^2.
\end{eqnarray}
Minimizing the free energy over the variational parameters $a$, $p$, and $q$ then gives the behavior near the transition, for $T<T_c$,
\begin{eqnarray}
a &=& \frac{K_{3}}{2\lambda}\left(\frac{\mu'}{K_{2}}\right)^{1/2}(T_c-T)^{1/2},\nonumber\\
p &=& \frac{K_{3}^2\mu'}{4\lambda^2}\left(\frac{3}{2 K_{2}\kappa}\right)^{1/2}(T_c-T),\\
q &=& \frac{1}{2}\left(\frac{3\mu'}{2\kappa}\right)^{1/2}(T_c-T)^{1/2}.\nonumber
\end{eqnarray}
Also, the free energy of the twist-bend phase just below the transition is
\begin{equation}
F_{TB}=-\frac{K_{3}^4 \mu'^3}{64 K_{2} \kappa \lambda^4}(T_c-T)^3.
\label{ftb}
\end{equation}

By comparison, to model the splay-bend phase, we make the variational ansatz inspired by the simulation results of Fig.~\ref{Fig9},
\begin{eqnarray}
\mathbf{\hat n}(x)&=&\cos\phi(x)\mathbf{\hat x}+\sin\phi(x)\mathbf{\hat z},\\
\mathbf{P}(x)&=&\frac{1}{2}p\cos(q x)\sin2\phi(x)\mathbf{\hat x}-p\cos(q x)\cos\phi(x)\mathbf{\hat z},\nonumber
\end{eqnarray}
where $\phi(x)=\alpha\sin(q x)$.
This ansatz also has three variational parameters:  $\alpha$ is the amplitude of the director modulation, $p$ is the amplitude of the polarization modulation, and $q$ is the modulation wavevector.  For this state, the splay, twist, and bend become
\begin{eqnarray}
\mathbf{S}(x)&=&-\frac{1}{2}q\alpha\cos(q x)\sin2\phi(x)\mathbf{\hat x}-q\alpha\cos(q x)\sin\phi(x)\mathbf{\hat z},\nonumber\\
\mathcal{T}(x)&=&0,\\
\mathbf{B}(x)&=&\frac{1}{2}q\alpha\cos(q x)\sin2\phi(x)\mathbf{\hat x}-q\alpha\cos(q x)\cos\phi(x)\mathbf{\hat z}.\nonumber
\end{eqnarray}
Here the twist is zero, as it should be for the splay-bend deformation.  The splay and bend both vary periodically through the modulated structure.  Plugging these quantities into Eq.~(\ref{fcomplete}) gives a free energy density that also varies periodically through the modulated structure.  We average the free energy density over the full modulation, and then minimize the average free energy over the variational parameters $\alpha$, $p$, and $q$.  This minimization gives the behavior near the transition,
\begin{eqnarray}
\alpha &=& \frac{K_{3}}{\lambda}\left(\frac{\mu'}{K_{1}}\right)^{1/2}(T_c-T)^{1/2},\nonumber\\
p &=& \frac{7 K_{3}^2\mu'}{8\lambda^2}\left(\frac{1}{2 K_{1}\kappa}\right)^{1/2}(T_c-T),\\
q &=& \frac{7}{8}\left(\frac{\mu'}{2\kappa}\right)^{1/2}(T_c-T)^{1/2}.\nonumber
\end{eqnarray}
Furthermore, the free energy of the splay-bend phase just below the transition is
\begin{equation}
F_{SB}=-\frac{K_{3}^4 \mu'^3}{32 K_{1} \kappa \lambda^4}(T_c-T)^3.
\label{fsb}
\end{equation}

From these results, we see that the uniform nematic phase can become unstable to the formation of \emph{either} the twist-bend phase or the splay-bend phase at the critical temperature $T_c$.  We can then ask which of these modulated phases is more stable.  Comparison of the free energies~(\ref{ftb}) and~(\ref{fsb}) shows that the twist-bend phase is more stable if $K_{1}>2K_{2}$, while the splay-bend phase is more stable if $K_{1}<2K_{2}$.  Interestingly, this is exactly the same criterion for the relative stability of the phases calculated by Dozov~\cite{dozov2001}.  This criterion is reasonable, because the elastic constants $K_{1}$ and $K_{2}$ give the energetic costs of splay and twist deformations, which are required in the splay-bend and twist-bend phases, respectively.

\section{Discussion}

In this paper, we have presented a theory for orientational order in bent-core liquid crystals.  The theory combines three parts: the Oseen-Frank free energy for director gradients, a tendency toward polar order perpendicular to the director, and a coupling between polar order and director bend.  In Landau theory, the Oseen-Frank free energy is represented by the $K$ terms, the tendency toward polar order by the $\mu$ term, and the coupling by the $\lambda$ term.  In the lattice model, the Oseen-Frank free energy is represented by $A$, the tendency toward polar order by $B_1$, and the coupling by $C$.  Either way, the physical conclusions are the same:  In the uniform nematic phase, there is a flexoelectric effect, where an imposed bend leads to a polarization, and conversely an applied electric field leads to a bend.  This flexoelectric effect increases as the temperature decreases toward a polar phase.  At the critical point, the response to an applied field diverges, and there is a second-order transition from the uniform nematic phase into a modulated polar phase.  The modulation may have the twist-bend or splay-bend structure, depending on the relative values of $K_1$ and $K_2$.

To compare our theory with the previous work by Dozov~\cite{dozov2001}, note that two phenomena occur at the critical temperature $T_c$:  the system transitions from the uniform nematic phase to the polar phase, and the renormalized bend elastic constant $K_{3}^\mathrm{eff}$ of Eq.~(\ref{K3eff}) changes sign from $K_{3}^\mathrm{eff}>0$ for $T>T_c$ to $K_{3}^\mathrm{eff}<0$ for $T<T_c$.  Dozov would say that the modulated phase is caused by the negative elastic constant $K_{3}^\mathrm{eff}$.  By contrast, we would say that the modulated phase \emph{and} the negative $K_{3}^\mathrm{eff}$ are both caused by the bend-polarization coupling together with the tendency toward polar order.  Of course, there is no contradiction between these two theories; they are just different ways of expressing the same physical concept.

Our theoretical results can be compared with recent experiments.  For the flexoelectric effect, the most relevant comparison is with the experiments of Harden \emph{et al.}~\cite{harden06, harden08}, which found a surprisingly large bend flexoelectric coefficient in bent-core liquid crystals, about three orders of magnitude larger than the typical value in rod-like liquid crystals.  This observation is at least qualitatively consistent with our concept that bent-core liquid crystals are near an incipient polar phase, and hence are very sensitive to any slight polar perturbations.  However, one aspect of this comparison is confusing.  In our theory, it is easy to understand why the polarization and bend induced by an applied electric field or an applied torque should be very large in bent-core liquid crystals, because these quantities diverge at the critical point.  By comparison, it is not easy to understand why the ratio between induced polarization and induced bend should be very large, because this ratio does not diverge at the critical point.  (The ratio increases as $T\to T_c$, but only toward a finite limit.)  Even so, that ratio is the standard definition of the flexoelectric coefficient, which Harden \emph{et al.}\ found to be surprisingly large.  One possible explanation of this discrepancy might be that the experiment involves local smectic order, which is not included in the theory; perhaps this smectic order increases the ratio of polarization to bend.  An alternative explanation might be that the experiment is somehow measuring one of the response coefficients that diverges at the critical point and is not actually measuring the ratio of polarization to bend.  Yet another possibility might be that the experiment is actually in a modulated polar phase that has not yet been identified.

For the modulated polar phases, the most relevant comparison is with recent experiments that find nonuniform nematic phases in systems of bimesogens~\cite{panov2010,luckhurst2011}.  These experiments provide good evidence that the observed modulation is a twist-bend structure, which is an encouraging consistency between theory and experiment.  However, one point is confusing in the comparison with Ref.~\cite{panov2010}.  The experiment reports that the periodicity of the observed stripe pattern is twice the thickness of the cell.  By contrast, the theory predicts that the periodicity should be determined by material parameters, even in the bulk liquid crystal, regardless of the cell thickness.  One possible explanation of this discrepancy might be that surface anchoring modifies the predictions of the theory and fixes the periodicity of the incipient modulation.  Such surface effects could be a topic for future research.

As a final point, we note that the modulated polar phases are locally ferroelectric; they have spontaneous polar order which leads to electrostatic polarization.  This local polarization is modulated in a helix for the twist-bend phase or a planar wave for the splay-bend phase, and hence it averages to zero globally.  For that reason, it might be difficult to observe in an experiment.  In this respect, the modulated polar phases are similar to ferroelectric smectic-C* liquid crystals, which \emph{also} have local polar order that is modulated in a helix and averages to zero globally.  By analogy with ferroelectric smectic-C* liquid crystals, there might be ways to unwind the helix of the twist-bend phase (or eliminate the wave modulation in a splay-bend phase) to obtain a structure with long-range polar order.  For example, strong surface anchoring might give a surface-stabilized ferroelectric nematic phase.  This surface stabilization could be another topic for future research.

\acknowledgments

We would like to thank D. W. Allender, A. J\'akli, R.~L.~B. Selinger, and T. J. Sluckin for helpful discussions.  This work was supported by the National Science Foundation through Grants DMR-0605889 and 1106014, and by the Samsung Electronics Corporation. Computational resources were provided by the Ohio Supercomputer Center and the Wright Center of Innovation for Advanced Data Management and Analysis.

\appendix*

\section{Mean-field calculation for lattice model}

The purpose of this Appendix is to show how the lattice model of Sec.~III can be solved analytically through mean-field theory, rather than Monte Carlo simulations.  The calculation is analogous to the mean-field theory in our previous paper~\cite{subas10}, but now for bend instead of splay flexoelectricity.  The results are similar to the simulations presented in Sec.~III.

\begin{figure}
\includegraphics[width=2.5in]{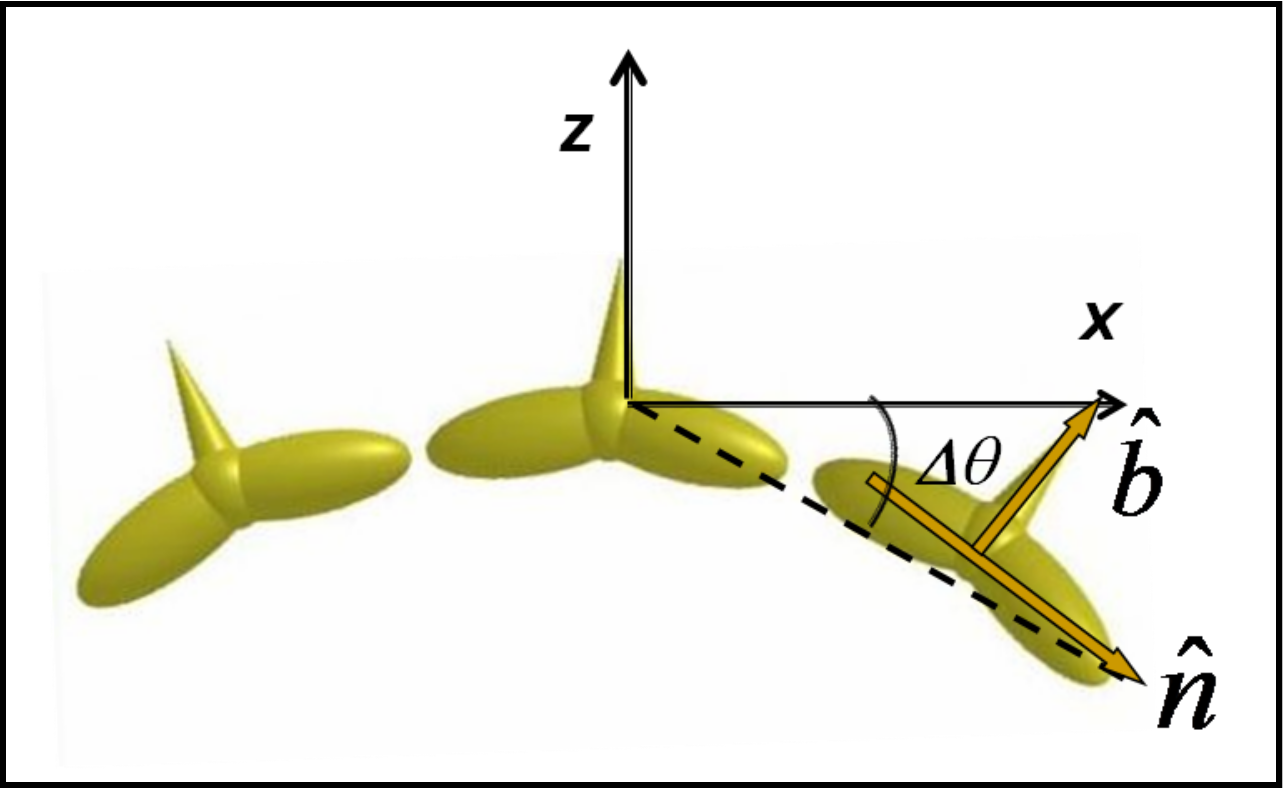}
\caption{\label{Fig10} (Color online) Construction for the lattice mean-field calculation.}
\end{figure}

For the mean-field calculation, we consider a small region of the system in the uniform nematic phase under an applied electric field.  We suppose that the field and average polarization are along the $\mathbf{\hat{z}}$ axis, and the director has a bend from layer to layer in the lattice as function of $x$, as shown in Fig.~\ref{Fig10}. Hence, we write
\begin{equation}
\mathbf{\hat{n}}=
\begin{cases}
(\cos\Delta\theta,0,-\sin\Delta\theta) &\mbox{in layer } 1,\\
(1,0,0) &\mbox{in layer } 2,\\
(\cos\Delta\theta,0,\sin\Delta\theta) &\mbox{in layer } 3.
\end{cases}
\end{equation}
Similarly, for the dipole direction $\mathbf{\hat{b}}_i$ on site $i$, we write
\begin{equation}
\mathbf{\hat{b}}_i=
\begin{cases}
(\sin\Delta\theta\cos\tilde{\theta}_i,\sin\tilde{\theta}_i,\cos\Delta\theta\cos\tilde{\theta}_i) &\mbox{in layer } 1,\\
(0,\sin\tilde{\theta}_i,\cos\tilde{\theta}_i) &\mbox{in layer } 2,\\
(-\sin\Delta\theta\cos\tilde{\theta}_i,\sin\tilde{\theta}_i,\cos\Delta\theta\cos\tilde{\theta}_i) &\mbox{in layer } 3.
\end{cases}
\end{equation}
We now suppose that the uniaxial nematic order is perfect, with a single well-defined value of the bend angle $\Delta\theta$, but there is a statistical distribution of the dipole orientations $\tilde{\theta}$.  In terms of this distribution, the polar order parameters is $P_1=\langle\cos\tilde{\theta}\rangle$ and the biaxial order parameter is $P_2=\langle\cos2\tilde{\theta}\rangle$.  If we average the lattice Hamiltonian of Eq.~(\ref{hamiltonian}) over the statistical distribution of $\tilde{\theta}$, and assume that $\Delta\theta$ is small, then we obtain the average energy per site
\begin{eqnarray}
U&=&-A\left[3-(\Delta\theta)^2\right] -B_1 P_1^2 \left[3 -\frac{1}{2}(\Delta\theta)^2\right]\nonumber\\
&&-B_2 \left[\frac{3}{2}(1+P_2^2) -\frac{1}{4}(1+P_2)^2(\Delta\theta)^2 \right] \label{appendixenergy}\\
&&-C P_1 \Delta\theta -E P_1\nonumber
\end{eqnarray}

\begin{figure}
\includegraphics[width=3.375in]{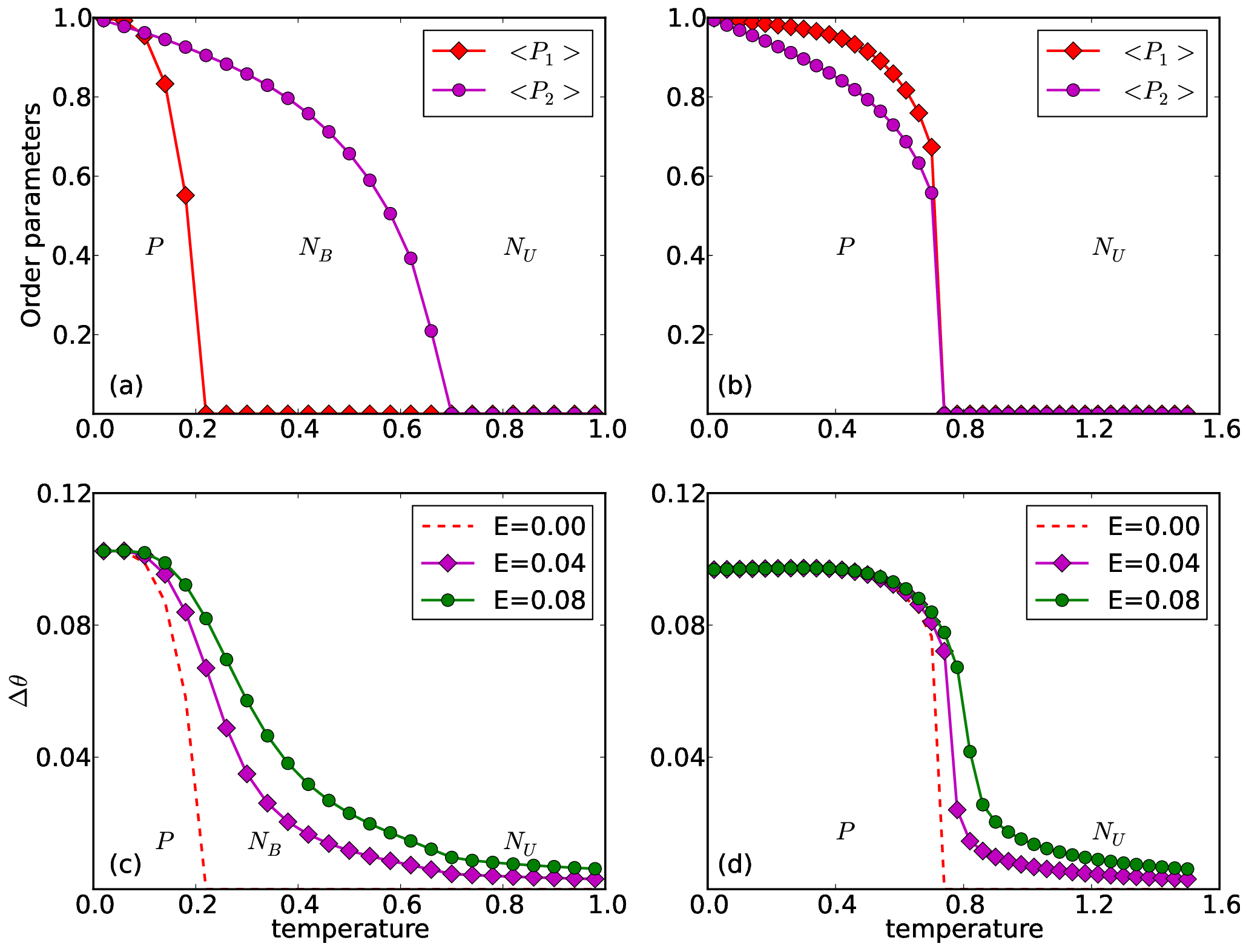}
\caption{\label{Fig11} (Color online) Mean-field results for the order parameters showing different types of transitions. (a, b) The biaxial order parameter $P_2$~($\bullet$) and polar order parameter $P_1$~($\diamond$) as functions of temperature $T$, for $B_1=0.03$ (in a) and  $B_1=0.20$ (in b) at zero field. (c, d) Bend $\Delta \theta$ as a function of temperature under an applied field. In all cases $A=1.0$, $B_2=0.45$, and  $C=0.30$.}
\end{figure}

To construct the free energy, we must combine the average energy with the entropy associated with the distribution of dipole orientations $\tilde{\theta}$.  The orientational distribution function is
\begin{equation}
\rho(\tilde{\theta})=\frac{e^{(v_1 \cos\tilde{\theta} + v_2 \cos2\tilde{\theta})}}%
{\int_0^{2\pi}e^{(v_1 \cos\tilde{\theta}+v_2 \cos2\tilde{\theta})} d\tilde{\theta}}
\end{equation}
where $v_1$ and $v_2$ are two parameters in the effective potential acting on $\tilde{\theta}$.  They are related to the order parameters $P_1$ and $P_2$ by
\begin{equation}
P_n(v_1,v_2)=\frac{\int_0^{2\pi} \cos(n\tilde{\theta}) e^{(v_1 \cos\tilde{\theta} + v_2 \cos2\tilde{\theta})}d\tilde{\theta}}%
{\int_0^{2\pi}e^{(v_1 \cos\tilde{\theta}+v_2 \cos2\tilde{\theta})} d\tilde{\theta}}
\end{equation}
for $n=1$ and $2$.
The entropic part of the free energy then becomes
\begin{eqnarray}
- T S &=& k_B T{\int_0^{2\pi}\rho(\tilde{\theta}}) \log(\rho(\tilde{\theta}))d\tilde{\theta}\nonumber\\
&=& k_B T\biggl[v_1 P_1 + v_2 P_2 \label{appendixentropy}\\
&&\qquad - \log\left(\int_0^{2\pi} e^{(v_1 \cos\tilde{\theta}+v_2 \cos2\tilde{\theta})}  d\tilde{\theta}\right)\biggr].\nonumber
\end{eqnarray}
Combining Eqs.~(\ref{appendixenergy}) and~(\ref{appendixentropy}), the full free energy is $F=U-TS$.

The problem to be solved is now:  For a given set of interaction parameters $A$, $B_1$, $B_2$, $C$, $E$, and temperature $T$, we should find the variational parameters $v_1$, $v_2$, and $\Delta\theta$ that minimize the free energy. Once we know the values of $v_1$ and $v_2$, we can calculate the order parameters $P_1$ and $P_2$. We can then determine whether the phase is uniaxial nematic ($N_U$), biaxial nematic ($N_B$), or polar ($P$), and we can see how the bend $\Delta\theta$ is related to the order parameters.

The free energy is minimized with Mathematica.  Figure~\ref{Fig11}a,b shows plots of the order parameters and the bend in zero field as functions of temperature.  Clearly the qualitative behavior of the phase transitions found in Monte Carlo simulations is reproduced. For small $B_1$ and $B_2$, there is a transition from $N_U \rightarrow N_B$, followed by a transition from $N_B \rightarrow P$ at lower temperature. For slightly larger $B_1$, the two transitions merge into a direct transition from $N_U \rightarrow P$. These mean-field results are generally consistent with the simulation results shown in Fig.~\ref{Fig4}, except that the transition temperatures are significantly higher. As mean-field theory exaggerates the tendency towards an ordered phase, it overestimates the transition temperatures. Despite this limitation, the simple mean-field calculation is successful in describing the qualitative behavior of the system.

To see the effect of an applied electric field, we repeat the minimization with a small field $E$ in the free energy.  Figure~\ref{Fig11}c,d shows the variation of the bend as a function of temperature under a field. These mean-field results are generally consistent with the simulation results shown in Fig.~\ref{Fig6}, showing a bend that responds more sensitively to any applied field as the temperature decreases toward the transition into the polar phase.

\end{document}